# Finite size effects in ferroelectric-semiconductor thin films under open-circuit electric boundary conditions

Eugene A. Eliseev[1], Sergei V. Kalinin[2], Anna N. Morozovska[3*]

[1]Institute of Problems for Material Sciences, NAS of Ukraine, 03028 Kyiv, Ukraine
[2]The Center for Nanophase Materials Sciences,
Oak Ridge National Laboratory, Oak Ridge, TN 37922, USA
[3]Institute of Physics, NAS of Ukraine, 03028 Kyiv, Ukraine

**Abstract**

General features of finite size effects in the ferroelectric-semiconductor film under open-circuit electric boundary conditions are analyzed using Landau-Ginzburg-Devonshire theory and continuum media electrostatics. The temperature dependence of the film critical thickness, spontaneous polarization and depolarization field profiles of the open-circuited films are found to be significantly different from the characteristics of short-circuited ones. In particular, we predict the re-entrant type transition boundary between the mono-domain and poly-domain ferroelectric states due to reduced internal screening efficiency and analyzed possible experimental scenarios created by this mechanism. Performed analysis is relevant for the quantitative description of free-standing ferroelectric films phase diagrams and polar properties. Also our results can be useful for the explanation of the scanning-probe microscopy experiments on free ferroelectric surfaces.

---

[*] Corresponding author: anna.n.morozovska@gmail.com

# I. Introduction

Ferroelectric thin films are actively explored in the context of fundamental physical studies and multiple applications [1, 2, 3]. In particular researchers are deeply interested in the determination of the film minimal (i.e. critical) thickness required for the development of ferroelectric instabilities, domain structure evolution including the thickness-induced transition into a single-domain state and, in general, the influence of finite size effect on ferroelectric properties [4].

One of the most important factors determining polarization and domain structure behavior in spatially confined ferroelectrics is the screening of bound charges related with spontaneous polarization divergence, div**P**, by external or internal free charge carriers. The discontinuity or/and divergence of polarization on the surface or/and in the bulk of ferroelectric means the presence of uncompensated bound charges which generate the internal electric field. Since the latter is pointed against the polarization in many cases, this field is called depolarization field. In the case of short circuited films the depolarization field is partially compensated by the screening charge in electrodes. For the insulating ferroelectric far from the sources of free charge carriers the depolarization field could totally destroy the polarization (unlike the ferromagnetic, where the demagnetization field energy is just the renomalization of the anisotropy energy). The internal screening in semiconducting ferroelectric leads to the restoring of polarization in the bulk of the thick films, but in the vicinity of surface the depolarization field is still strong and polarization should drop almost to zero here and thus is weakly dependent on the surface energy (i.e. on the extrapolation length lambda). The screening mechanism defines the depolarization electric field structure and value, which in turn determines the critical thickness and other size effects [5, 6, 7].

The screening of the ferroelectric polarization oriented normally to the film surface plane (further regarded as *out-of-plane* polarization) in thin film covered by perfect electrodes is mainly external and the semiconductor properties of the material (if any) do not play any essential role in the screening mechanism. In the presence of chemically active species, the effective screening is induced by electrochemical reactions on ferroelectric surface with charged ionic species bounded to polarization charges [8, 9, 10, 11]. However, the rate of these chemical screening is limited by mass transport and requires availability of screening species.

Here, we consider the case of the polarization screening in the film without one or both electrodes and cleaned free surface placed in dielectric ambient. This corresponds to the material rapidly cooled from above Curie temperature, material in the ultra-high vacuum environment, or freshly formed cleave or break surface. In this case, the screening can be internal, i.e. performed by the free carriers inside the film. These most common realistic situations correspond to the limiting cases of short- or open-circuit electric boundary conditions shown in the **Figure 1**. Polar axis lies normal to the film surface plane.

Finite size effects in single- and poly-domain ferroelectric films under short-circuit conditions are studied in details both numerically and analytically [1].

In the beginning Onsager [12] and Landau [13] predicted theoretically the existence of free-standing two-dimensional ferroelectrics. In their pioneer work Bune et al [14] proved experimentally that ferroelectric state is possible for one-monolayer thick dielectric PVDF films. Later on Fong et al [15] and Chanthabouala et al [16] revealed that several unit cell thick $PbTiO_3$ and $BaTiO_3$ films can be ferroelectric when covered by conducting electrodes and epitaxially clamped to perovskite substrates, which possible role is to create epitaxial strain, that is strong enough to maintain the ferroelectric phase via electrostriction mechanism [17]. From the first-principles calculations, Junquera and Ghosez predicted that $BaTiO_3$ thin films clamped between two metallic $SrRuO_3$ electrodes in short circuit lose their ferroelectric properties below a critical thickness of about six unit cells (~24 Å) [18]. Using first-principles calculations and phenomenological Landau-Ginzburg-Devonshire (LGD) theory for comparison, Duan et al [19] have shown that calculated from the first-principles critical thickness in $KNbO_3$ with Pt electrodes is about 1nm, while chosen phenomenological parameters do not allow to reproduce qualitatively the first-principles results. As one can conclude from the experimental results [14, 15, 16, 18, 19] the critical thickness of ferroelectricity disappearance can be very small or even absent for the short-circuited boundary conditions, since the conditions minimize depolarization field (if any), in contrast to the open-circuited ones. Note, that LGD theory predictions can be valid for thicknesses much higher than a lattice constant, that is possibly true for the critical thickness of the open circuit films that is expected to vary from tens of nanometers to tens of micrometers.

The single-domain ferroelectric-semiconductor film under open-circuit conditions was firstly considered by different authors [20, 21, 22, 23, 24] long ago, but the analytical study of finite size effects is absent to date, and, in particular, no phase diagrams and analytical expressions for the critical thickness were available. Since the open-circuit conditions correspond to the conventional scanning-probe microscopy (SPM) geometry used for the modern investigation of ferroelectric-semiconductor films (such as $BaTiO_3$ (BTO) [25], $BiFeO_3$ (BFO) [26, 27], and $PbZr_xTi_{1-x}O_3$ (PZT) [28, 29, 30]), we decided to revisit the problem.

We further note that this mechanism is of interest for materials with sufficiently high carrier concentration only, i.e. strong chemical doping or small band gap ferroelectrics. Indeed when the concentration of free carriers becomes lower, critical thickness becomes higher, and eventually the film will split into domains in order to decrease positive depolarization field energy. The scenario is realized because the screening by free carriers decreases the depolarization field energy and does not change the domain wall energy. The critical concentration of carriers required for the domain splitting is discussed by the Tagantsev and Fousek (see [31] and refs. therein [32, 33, 34]). For sufficiently high concentrations the analysis of a single-domain state in the open-circuited ferroelectric-semiconductor film is physically reasonable and presented below.

## II. Basic equations

Within the framework of the LGD theory, equilibrium one-dimensional distribution of the polarization component $P_3(z)$ in a single-domain ferroelectric film can be found from the Euler-Lagrange equation with boundary conditions:

$$\begin{cases} \alpha P_3 + \beta P_3^3 + \gamma P_3^5 - g_{33} \frac{\partial^2 P_3}{\partial z^2} = -\frac{\partial \varphi}{\partial z}, \\ \left( P_3 + \lambda \frac{\partial P_3}{\partial z} \right) \bigg|_{z=+L/2} = 0, \quad \left( P_3 - \lambda \frac{\partial P_3}{\partial z} \right) \bigg|_{z=-L/2} = 0. \end{cases} \quad (1)$$

Here $\alpha$, $\beta$ and $\gamma$ is the expansion coefficients of LGD model; $g_{33}$ is the component of tensor of gradient energy coefficients, $\lambda$ is the extrapolation length [5, 35]. Electric potential $\varphi$ can be found self-consistently from the Poisson equation with the open-circuit electric boundary conditions:

$$\begin{cases} \varepsilon_{33}^b \frac{\partial^2 \varphi}{\partial z^2} = \frac{1}{\varepsilon_0} \left( \frac{\partial P_3}{\partial z} - \rho(\varphi) \right) \\ D_3 \big|_{\pm L/2} \equiv \left( -\varepsilon_0 \varepsilon_{33}^b \frac{\partial \varphi}{\partial z} + P_3 \right) \bigg|_{\pm L/2} = 0 \end{cases}, \quad (2)$$

Here $\varepsilon_{33}^b$ is the background permittivity of ferroelectric [36], $L$ is the film thickness (see **Figure 1**). Boundary conditions in Eq.(2) are used by many authors [37, 38, 39] for the "clean" non-conducting interfaces. They are based on several assumptions, namely (a) free surface charge is absent, so that electrical displacement should be continues at the interface; (b) at the same time electric field outside the crystal should be set to zero, otherwise the energy of the system per unit area may reach infinity.

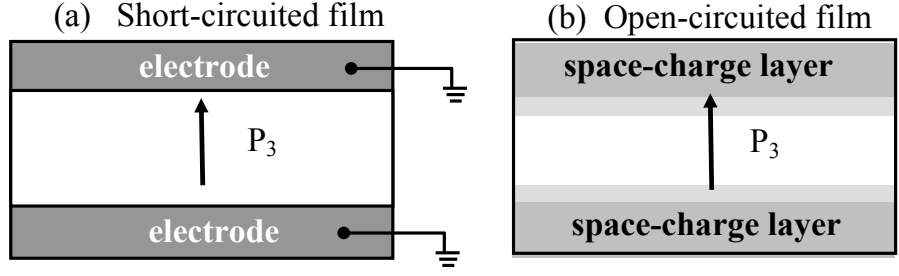

**Figure 1.** Schematics of the single-domain **(a)** short-circuited ferroelectric-dielectric film, **(b)** open-circuited ferroelectric-semiconductor film. Polarization direction is shown by arrow.

In was shown that the electronic gas is degenerated in the vicinity of domain walls in typical ferroelectric ilmenites [40] and perovskites [41, 42]. So it is naturally to assume that the degeneration is probably in the vicinity of the film surfaces, when the film is far from the phase transition from ferroelectric to paraelectric phase. Here the Fermi-Dirac statistics should be used for the charge density in the Poisson equation [43]. For a donor-doped semiconductor it can be modeled as [40]:

$$\rho(\varphi) = N_d^+(\varphi) - n(\varphi) \quad (3a)$$

The donor level is regarded infinitely thin with activation energy $E_d$. For the case the concentration of donors is determined by a single Fermi-Dirac distribution function [44]:

$$N_d^+(\varphi) = N_{d0}(1 - f(E_d - E_F - q\varphi)), \quad (3b)$$

$N_{d0}$ is the concentration of donor centers, $f(x) = \dfrac{1}{1 + \exp(x/k_B T)}$ is the Fermi-Dirac distribution function, $k_B = 1.3807 \times 10^{-23}$ J/K, $T$ is the absolute temperature. $E_F$ is the Fermi energy level, $E_d$ is the donor level (all energies are counted from the vacuum level). The concentration of the electrons in the conductive band considered in the continuous levels approximation [45, 46] is:

$$n(\varphi) = \int_0^\infty d\varepsilon \cdot g_n(\varepsilon) f(\varepsilon + E_C - E_F - q\varphi). \quad (3c)$$

$E_C$ is the bottom of conductive band. The "bulk" density of states in the effective mass approximation is $g_n(\varepsilon) \approx \dfrac{\sqrt{2 m_n^3 \varepsilon}}{2\pi^2 \hbar^3}$, one obtains the approximate equalities in Eq.(3c),

$$n(\varphi) \approx \left(\frac{m_n k_B T}{\hbar^2}\right)^{3/2} \frac{1}{\pi^2 \sqrt{2}} \frac{\sqrt{\pi}}{2}\left(-\mathrm{Li}_{3/2}\left(-\exp\left(\frac{q\varphi + E_F - E_C}{k_B T}\right)\right)\right), \text{ where } \mathrm{Li}_n(z) = \sum_{k=1}^{\infty} \frac{z^k}{k^n} \text{ is the}$$

polylogarithmic function. Fermi level position is determined in a self-consistent way from the electroneutrality condition $\rho(0) = N_{d0}^+ - n_0 = 0$, where $N_{d0}^+ = N_{d0} f(E_F - E_d)$ and

$$n_0 = \int_0^\infty d\varepsilon \cdot g_n(\varepsilon) f(\varepsilon + E_C - F_F)$$

### III. The film critical thickness

In order to derive analytical expressions for the film critical thickness corresponding to the second order transition into the paraelectric phase, one can use Debye approximation for the charge density (3) in the immediate vicinity of the transition, since the spontaneous polarization and corresponding depolarization field become very small here. In accordance with our numerical calculations based on Eqs.(3) Debye approximation is valid in the immediate vicinity of the second order phase transition into the paraelectric phase from either single-domain or poly-domain ferroelectric states, because the condition of its validity, $|e\varphi/k_B T| \ll 1$, holds true here.

#### 3.1. The transition from paraelectric to a single-domain ferroelectric state

In Debye approximation, for the case of the screening by non-degenerated free carriers (i.e. when $|e\varphi/k_B T| \ll 1$), 1D Poisson Eq.(2) acquires the form, $\dfrac{\partial^2 \varphi}{\partial z^2} - \dfrac{\varphi}{R_d^2} = \dfrac{1}{\varepsilon_0 \varepsilon_{33}^b} \dfrac{\partial P_3}{\partial z}$, where $R_d = \sqrt{\varepsilon_0 \varepsilon_{33}^b k_B T / (2 e^2 n)}$ is Debye screening radius. For numerical calculations we will assume that the temperature dependence of the carries concentration $n$ obeys the activation law, $n \approx N_0 \exp(-E_d/k_B T)$ [43]. Note, that the temperature dependence of Debye screening radius is almost indifferent on ferroelectric material parameters, it depends on $n_0$ and $\varepsilon_{33}^b$ only. The 1D approximation is valid in a single-domain ferroelectric and paraelectric phases of the film.

In the Debye approximation the potential and depolarization field in the material are given by expressions:

$$\varphi(z) = \frac{R_d P_3(L/2)}{\varepsilon_0 \varepsilon_{33}^b} \frac{\sinh(z/R_d)}{\sinh(L/2R_d)} - \frac{1}{\varepsilon_0 \varepsilon_{33}^b} \int_{-L/2}^{L/2} \left(\frac{dP_3}{d\xi}\right) G(\tilde{z}, z) d\tilde{z} \qquad (4a)$$

$$E_3^d(z) = -\frac{\partial \varphi}{\partial z} = -\frac{P_3(L/2)}{\varepsilon_0 \varepsilon_{33}^b} \frac{\cosh(z/R_d)}{\sinh(L/2R_d)} + \frac{1}{\varepsilon_0 \varepsilon_{33}^b} \int_{-L/2}^{L/2} \left(\frac{dP_3}{d\tilde{z}}\right) \frac{\partial G(\tilde{z}, z)}{\partial z} d\tilde{z} \qquad (4b)$$

Allowing for the symmetry of boundary problem Eq. (1,2), the spontaneous polarization distribution is symmetric with respect to the film center, i.e. $P_3(L/2) = P_3(-L/2)$. In this case, the Green's function is

$$G(\tilde{z}, z) = -\frac{R_d}{2} \left( \exp\left(-\frac{|\tilde{z}-z|}{R_d}\right) + \frac{\exp\left(-\frac{L/2+\tilde{z}}{R_d}\right)\cosh\left(\frac{z-L/2}{R_d}\right) + \exp\left(-\frac{L/2-\tilde{z}}{R_d}\right)\cosh\left(\frac{z+L/2}{R_d}\right)}{\sinh(L/R_d)} \right)$$

(4c)

The critical thickness of the film transition into a paraelectric phase can be found from the characteristic equation obtained from the boundary problem (1) and (2) by differentiation, namely from the system of equations listed in Ref.[47]. It can be shown that the critical thickness corresponds to the lowest solution of the approximate transcendental equation:

$$\tan\left(\frac{L}{2\kappa}\right) = \pm\left(\frac{\kappa}{\lambda}\frac{g_{33}}{\alpha}\left(\frac{1}{\kappa^2}+\frac{1}{\xi^2}\right)\left(1+\frac{\lambda}{\xi}\right) - \frac{\kappa}{\xi}\right), \tag{4a}$$

namely:

$$L_{cr} = 2\kappa \arctan\left(-\frac{g_{33}}{\alpha}\frac{\kappa}{\lambda}\left(\frac{1}{\kappa^2}+\frac{1}{\xi^2}\right)\left(1+\frac{\lambda}{\xi}\right)+\frac{\kappa}{\xi}\right) \approx \pi\kappa. \tag{4b}$$

Here the screening length $\kappa$ and the longitudinal correlation length $\xi$ are introduced. They are given by expressions $\kappa \approx R_d / \sqrt{-\varepsilon_{33}^b \varepsilon_0 \alpha}$ and $\xi = \sqrt{\varepsilon_{33}^b \varepsilon_0 g_{33}}$ correspondingly. The screening length can be rather high (i.e. the order of 10 $R_d$), while the correlation length is typically smaller than the lattice constant due to depolarization effects [5, 31]. The equations (4) are valid under the conditions $|e\varphi/k_B T| \ll 1$, $L \gg \xi$ and have physical roots at $\alpha < 0$. The coefficient $\alpha = \alpha_T(T - T_c^*)$ for the strained ferroelectric film, $T_c^* = T_c + \frac{u_m^*}{\alpha_T}\frac{4Q_{12}}{s_{11}+s_{12}}$ is the Curie temperature renormalized by the epitaxial misfit strain $u_m = (a/c) - 1$.

### 3.1. The transition from paraelectric to a poly-domain ferroelectric state

The internal screening competes with domain formation as the mechanism of depolarization energy minimization for small carrier concentrations. Tagantsev and Fousek [31] underlined that all available theories [32-34] predict an increase in the domain period with increasing of free carriers concentration, followed by a transition to a single-domain state a high enough level of screening. In particular, the period of the domain pattern $W$ can deviate from the Kittel-type law, $W \propto \sqrt{L}$, due to the screening by free carriers ($L$ is the film thickness).

We have shown (see Supplementary Materials [47]) that the transition between paraelectric and multidomain ferroelectric state takes place at the critical film thickness

$$L \approx \frac{2\pi}{\sqrt{\varepsilon_{33}^b \varepsilon_0 g_{55}}} \left( -\frac{\alpha}{g_{55}} + \frac{\varepsilon_{33}^b}{\varepsilon_{11}^b R_d^2} \right)^{-1}.$$ (5a)

Here $g_{55}$ is the component of tensor of gradient energy coefficient, in general case is different from component $g_{33}$ introduced earlier, At the transition the emerging domain structure could be characterized with the wave vector

$$k_{min} = \sqrt{-\frac{\alpha}{2g_{55}} - \frac{\varepsilon_{33}^b}{2\varepsilon_{11}^b R_d^2}} \equiv \sqrt{\frac{\pi}{L\sqrt{\varepsilon_{33}^b \varepsilon_0 g_{55}}} - \frac{\varepsilon_{33}^b}{\varepsilon_{11}^b R_d^2}}$$ (5b)

From Equation (5b) the screening radius decrease leads to the decrease of wave vector and finally to the disappearance of domains in the point where the following condition is true $-\alpha = \frac{\varepsilon_{33}^b}{\varepsilon_{11}^b R_d^2} g_{55}$. According to Eq.(5b) the increase of carriers concentration leads to the increase of domain structure period at the transition between paraelectric and poly-domain ferroelectric state.

Since free carriers concentration in the ferroelectrics-semiconductors can readily vary in the range $n_{cr} = (10^{25} - 10^{23})$ m$^{-3}$ the single-domain state is energetically preferable for films with thickness more than 10-100 nm under the open-circuit conditions. Hence, below we estimate the critical concentration for the onset of domain splitting and analyze finite size effects and phase diagrams of single-domain ferroelectric-semiconductor film under these conditions.

### IV. Numerical results and their analyses

As shown in the **Figures 2a-d,** the critical thickness (4b) can vary in a wide range, from tens of nanometers to tens of micrometers, depending on the ferroelectric material parameters, temperature and carriers concentration. Note that the critical thickness (4b) is almost independent on the extrapolation lengths, i.e. the approximation $L_{cr} \approx \pi\kappa$ is rather rigorous for the open-circuit conditions.

Hence, $L_{cr} \propto \pi R_d / \sqrt{-\varepsilon_{33}^b \varepsilon_0 \alpha}$ in general case. The thickness (4b) can be compared with the value $\widetilde{L}_{cr} \approx -\frac{2\xi^2}{\varepsilon_0 \alpha \varepsilon_{33}^b (\xi + \lambda)}$ calculated for a short-circuited film. As anticipated $\widetilde{L}_{cr} \ll L_{cr}$ entire all reasonable range of material parameters, suggesting the reduction of ferroelectric phase stability in the open-circuited film.

At fixed film thickness $L$ the critical temperature, $T_{cr}(L)$, can be found as a solution of transcendental equation $L \approx \pi\kappa(T_{cr})$, or in explicit form:

$$T_{cr} = T_c^* - \frac{1}{\varepsilon_{33}^b \varepsilon_0 \alpha_T}\left(\frac{\pi R_d(T_{cr})}{L}\right)^2 \qquad (6)$$

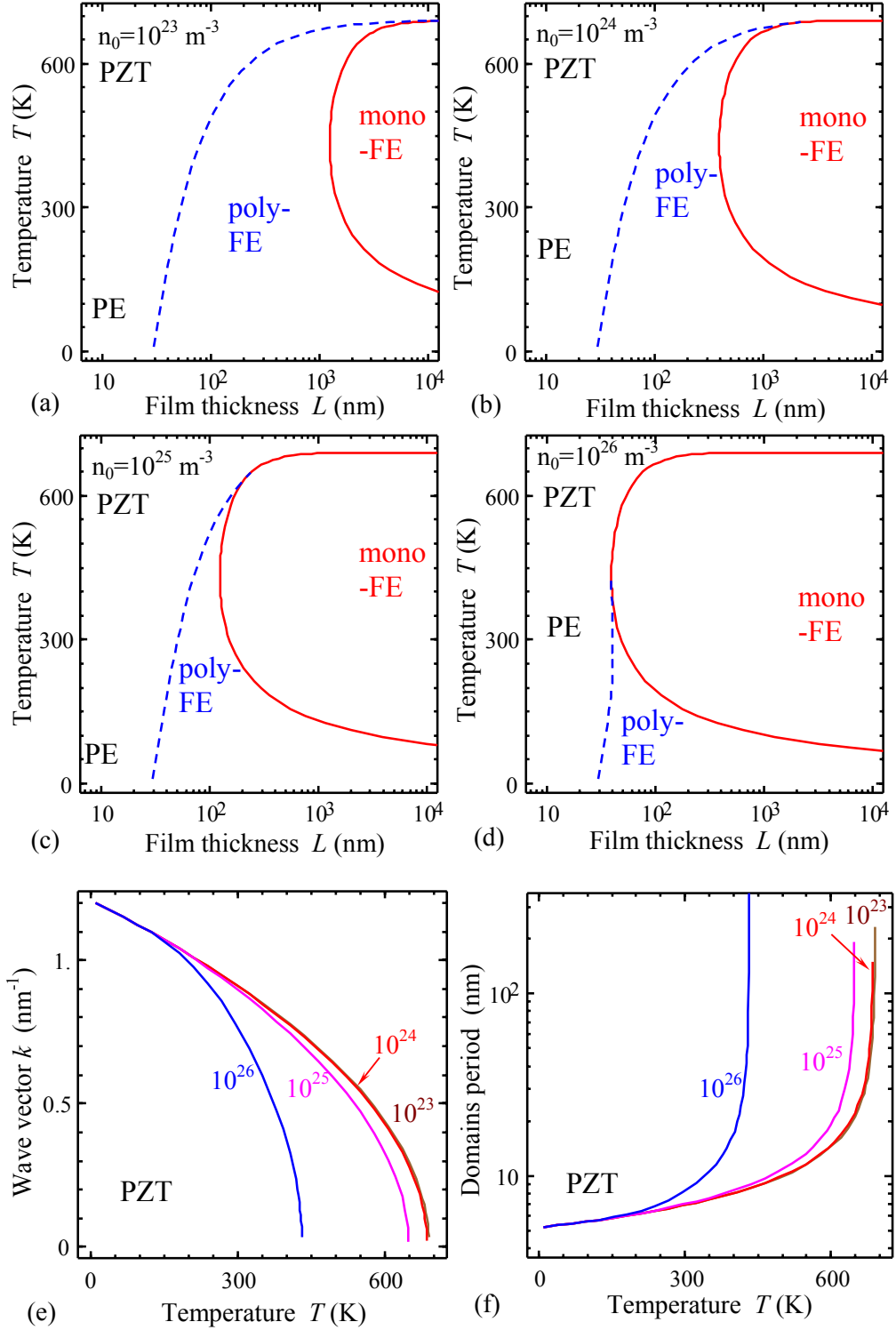

**Figure 2.** Phase diagram in coordinates "film thickness – temperature" calculated for PbZr$_{40}$Ti$_{60}$O$_3$ (PZT) films for different concentrations $n_0$=10$^{23}$, 10$^{24}$, 10$^{25}$ and 10$^{26}$ m$^{-3}$ (plots (a), (b), (c) and (d) respectively). Temperature dependence of wave vector (e) and corresponding period (f) of domain structure for different concentrations $n_0$=10$^{23}$, 10$^{24}$, 10$^{25}$ and 10$^{26}$ m$^{-3}$

(numbers near the curves) Abbreviations PE, mono-FE and poly-FE denote paraelectric, mono-domain and poly-domain ferroelectric phase regions correspondingly.

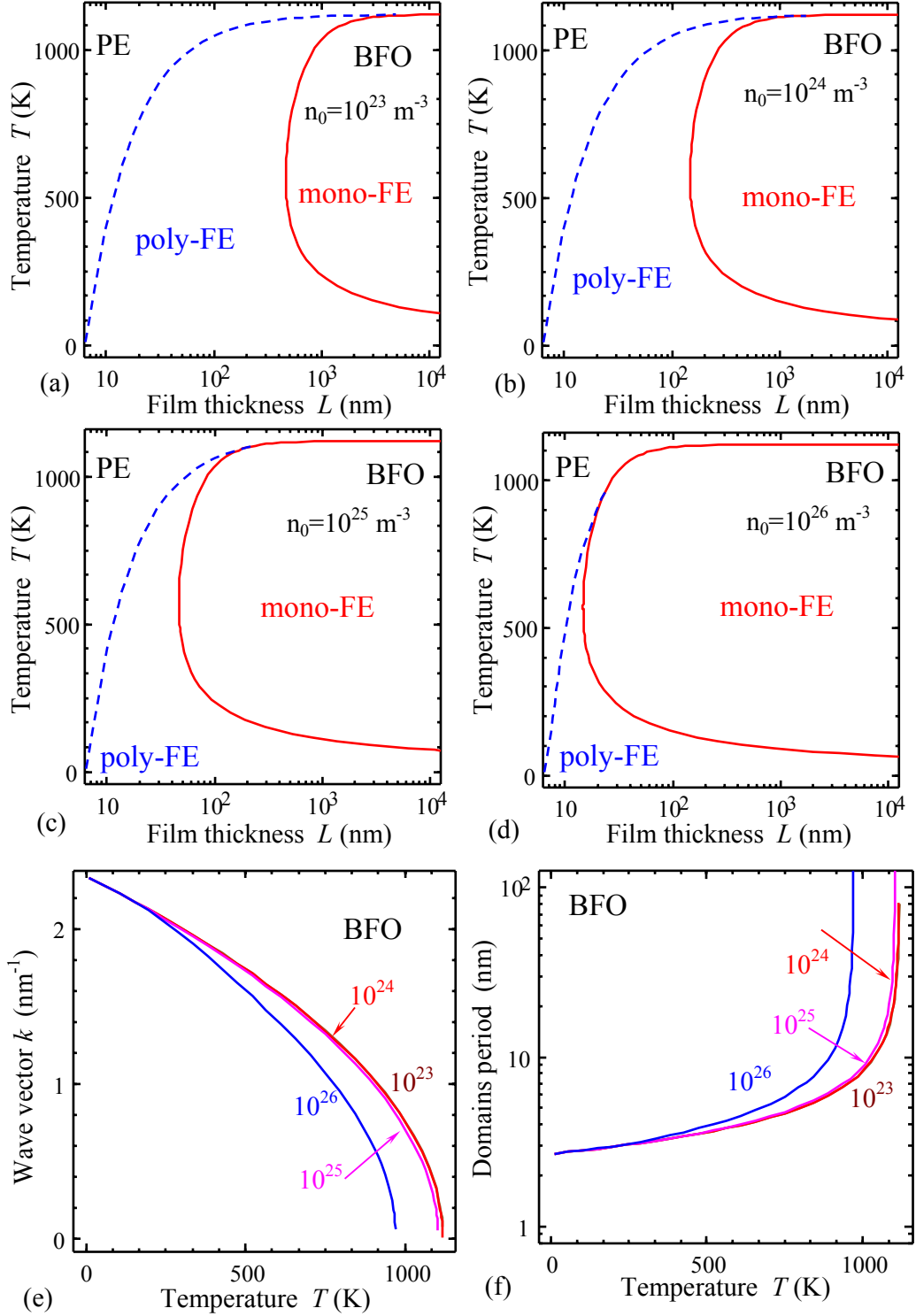

**Figure 3.** Phase diagram in coordinates "film thickness – temperature" calculated for BFO films for different concentrations $n_0 = 10^{23}$, $10^{24}$, $10^{25}$ and $10^{26}$ m$^{-3}$ (panels (a), (b), (c) and (d) respectively). Temperature dependence of wave vector (e) and corresponding period (f) of domain structure for different concentrations $n_0 = 10^{23}$, $10^{24}$, $10^{25}$ and $10^{26}$ m$^{-3}$ (numbers near the curves) Abbreviations PE, mono-FE and poly-FE denote paraelectric, monodomain and polydomain ferroelectric phase regions correspondingly.

Note that the **Figures 2** and **3** can be interpreted as a phase diagram in coordinates "film thickness – temperature". The unusual "re-entrant" type shape of monodomain ferroelectric – paraelectric phase transition boundary (solid curve) originated from the semiconductor properties contribution, namely from the temperature dependence of the Debye screening radius $R_d$. Further the transition to poly-domain state (dashed curve) occurs with either temperature or film thickness decrease.

Table 1. Material parameters for bulk ferroelectrics-semiconductors

| coefficient | BiFeO$_3$ [48]) | PbZr$_{40}$Ti$_{60}$O$_3$ [49] |
|---|---|---|
| $\varepsilon_{33}^b$ | 9 | 6 |
| $\alpha$ ($\times 10^7$ C$^{-2}\cdot$mJ) at 293°K | −8.1046 | −16.88 |
| $\alpha_T$(C$^{-2}\cdot$mJ/K) | 9.8×10$^5$ | 4.24×10$^5$ |
| $T_C$ (K) | $T_C$=1120 | 691 |
| $\beta$ (C$^{-4}\cdot$m$^5$J) | 26×10$^8$ | 1.445×10$^8$ |
| $\gamma$ (C$^{-6}\cdot$m$^9$J) | 0 | 1.14×10$^9$ |
| $g$ ($\times 10^{-10}$ C$^{-2}$m$^3$J) | 1 | 1 |
| $n_0$ (m$^{-3}$) | 10$^{25}$ | 10$^{25}$ |
| $E_d$ (eV) | 0.1 | 0.1 |

We further analyze the spontaneous polarization distribution and amplitude at film thickness $L > L_{cr}$ using Fermi-Dirac statistics for electrons and donors. Note, that the curves in the **Figure 4** were calculated numerically beyond the limit of Debye approximation, i.e. using the nonlinear carrier density dependence on the potential taking into account degeneration of carriers (Eqs.(3b,c)).

Polarization and depolarization field values and z-profiles shape strongly depends on the concentration $n_0$ (see **Figure 4**); meanwhile they appeared almost independent on extrapolation lengths. Polarization value increases and tends to the bulk value in the central part of the film; its profile becomes more flat with $n_0$ increase (**Figures 4a,b**). Depolarization field value in the central part of the film vanishes; its profile stops oscillating and becomes more flat with $n_0$ increase (**Figures 4c,d**). As anticipated the profiles tends to the homogeneous distribution in the central part of the film with its thickness increase.

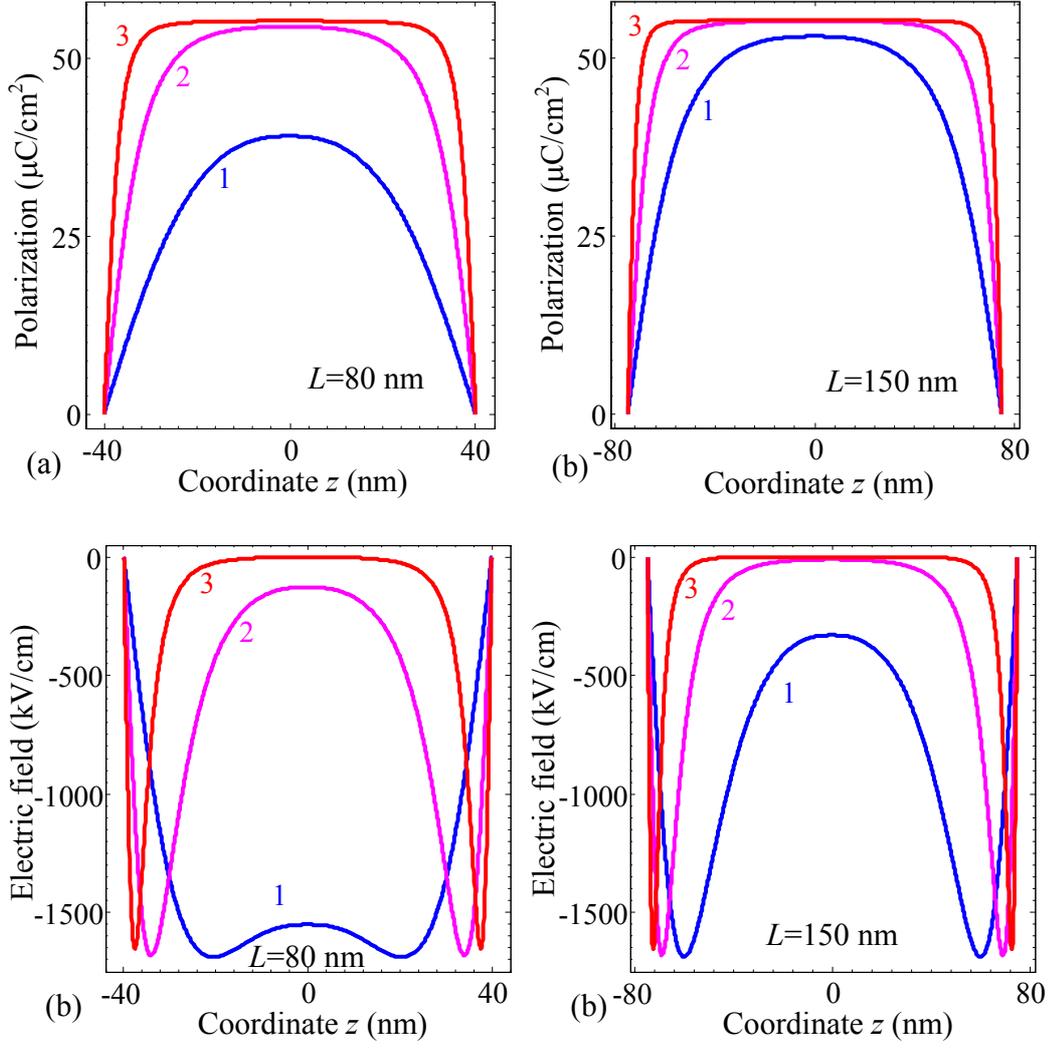

**Figure 4.** Polarization **(a,b)** and depolarization field **(c,b)** profiles in the 80 and 150 nm - thick BFO film calculated at room temperature for different concentration $n_0 = 10^{24}$, $10^{25}$ and $10^{26}$ m$^{-3}$ (curves 1, 2 and 3 correspondingly).

Despite the polarization distribution has only one length-scale and polarization value at the surface strongly depends on the so-called extrapolation length lambda for short-circuited ferroelectric films, for open-circuited films we indeed obtained that the depolarization field has a drastic effect on the polarization profile and its surface value. Since the depolarization field is unscreened near the surface, it suppresses both polarization and its derivative almost to zero (so the boundary condition in Eq.(1) remains valid), while deeper into the film, namely at distances higher than a few screening length, depolarization field exponentially decreases and polarization restores to its bulk value. That is why the critical thickness for open-circuited film is simply proportional to screening length with great accuracy and almost independent on the extrapolation length. As for the polarization profile near the surface, here the longer length scale (screening radius) dominates over the smaller one (correlation length) and the polarization distribution is almost independent on the extrapolation length (some dependence could be seen near the surface at the layer with thickness of about few correlation lengths). One could see the effects of extrapolation length on polarization profile only for the very high concentration of carriers when the screening radius decreases (see **Figure 5**).

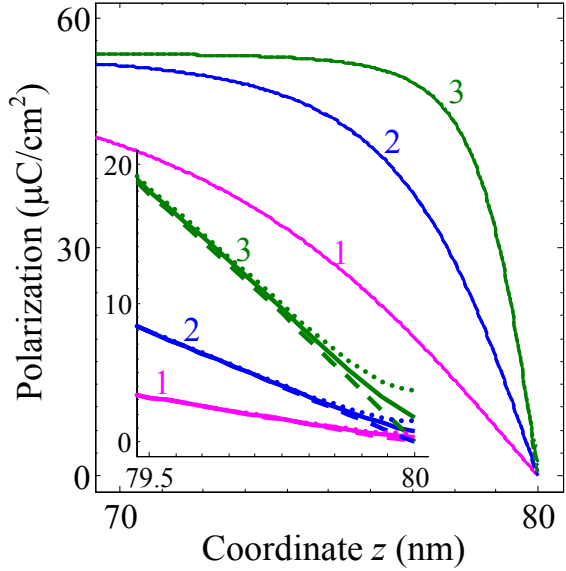

**Figure 5.** Polarization profile in the open-circuit film of 100 nm thickness for different concentration of screening carriers equal to $10^{25}$, $10^{26}$, $10^{27}$ m$^{-3}$ (curves 1, 2 and 3) and extrapolation length values $\lambda=0$, 0.3 and 100 nm (dashed, solid and dotted curves). Inset shows polarization behavior near the surface.

### IV. Discussion

Finally, we would like to underline that the critical thickness value, phase diagrams and profiles calculated for the ferroelectric-semiconductor film under open-circuit conditions exhibit strong differences in comparison with the well-studied case of ferroelectric-dielectric film under short-circuit conditions. Based on the analytical results Eqs.(4)-(6) **Figures 6** schematically illustrate the general characteristics of these differences. For the case of short-circuited film the transition temperature $T_{cr}$ into ferroelectric phase depends of the film thickness $L$ monotonically; its starts at critical thickness $L_{cr}$, increases with the film thickness and saturates to the bulk value (see dashed curve in **Figures 6a**). For the open-circuited films we predict the re-entrant type transition boundary between the mono-domain and poly-domain ferroelectric states (see solid curve in **Figures 6a**). This suggests the possibility of the re-entrant transition for the rapidly cooled or freshly prepared ferroelectric surfaces and thin ferroelectric films. The overall behavior of the system will then be controlled by the relative rates and efficiency of internal vs. external screening processes. For example, slow cooling of ferroelectric leads to the formation of the classical ferroelectric phase stabilized by internal screening, which is slowly replaced by mass-rate limited external screening and persists on subsequent cooling. At the same time, rapid cooling of ferroelectric can lead to the situation when polarization is "overcooled" and will appear only due to the external screening processes if allowed by the composition of the system.

We further note that this behavior can affect polarization switching in the local probe based experiments. Here, application of tip bias and surface electrochemical processes can suppress polarization and induce local paraelectric state, rather than induce polarization switching. Full analysis of this behavior then requires consideration of detailed thermodynamics of chemical screening.

There are further significant differences in the spontaneous polarization and depolarization field profiles for the film thickness above the critical one. Polarization profile is typically smooth for the open-circuited film and saturated for the short-circuited one (compare solid and dashed curves in **Figures 6b**). Depolarization field profile in the open-circuited film is non-monotonic with three local maxima separated by two minima (solid curve in **Figures 6c**), while its profile is almost constant inside the film and strongly increases only in the immediate vicinity of the surfaces of the short-circuited film (dashed curve in **Figures 6c**). These behaviors can be anticipated to be visible in surface-sensitive techniques, e.g. X-ray reflectometry and electron microscopy of ferroelectrics edges.

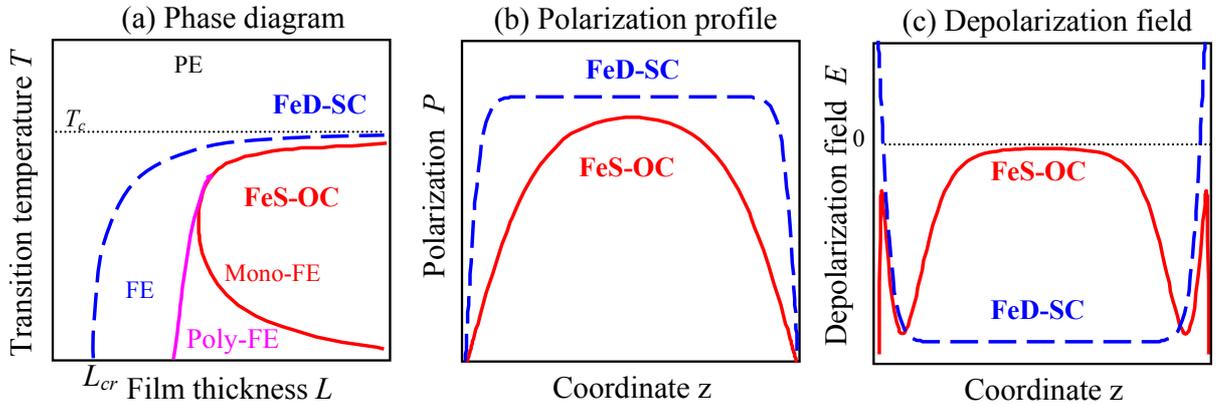

**Figure 6. (a)** Phase diagram in coordinates temperature - film thickness. Polarization **(b)** and depolarization field **(c)** profiles. Abbreviation **FeS-OC** denotes ferroelectric-semiconductor film under open-circuited electric boundary conditions (solid curves); **FeD-SC** denotes ferroelectric-dielectric film under short-circuited electric boundary conditions (dashed curves).

## V. Conclusion

To summarize, we derived analytical expressions, which allow establishing the general features of finite size effects in the ferroelectric-semiconductor film under open-circuit electric boundary conditions. There are essential differences in the temperature dependence of the film critical thickness, spontaneous polarization and depolarization field profiles of the open-circuited films in comparison with the short-circuited ones. Since the electric boundary conditions corresponding to the conventional SPM geometry can be regarded open-circuit, calculations of the critical thickness for different ferroelectric-semiconductor films give us the information that can be of practical importance.

## Acknowledgements

Authors are very grateful to the constructive critical remarks and stimulating discussions from Prof. A.K. Tagantsev. E.A.E. and A.N.M. acknowledge Center for Nanophase Materials Sciences (CNMS), user projects CNMS 2013-293 and CNMS 2014-270, and National Academy of Sciences of Ukraine (grant 35-02-14). S.V.K. acknowledges Office of Basic Energy Sciences, U.S. Department of Energy.

**References**

[1] D.G. Schlom, L.-Q. Chen, Ch.-B. Eom, K.M. Rabe, S.K. Streiffer, and J.-M. Triscone. Annu. Rev. Mater. Res. 37:589–626 (2007).

[2] A.L. Kholkin, S.V. Kalinin, A. Roelofs, and A. Gruverman. "Review of ferroelectric domain imaging by piezoresponse force microscopy." In Scanning Probe Microscopy, pp. 173-214. Springer New York, 2007.

[3] J. F. Scott, Nature materials 6, no. 4: 256-257 (2007).

[4] D.R. Tilley. Finite-size effects on phase transitions in ferroelectrics, Ferroelectic Thin Films. ed. C. Paz de Araujo, J.F.Scott and G.W. Teylor. Amsterdam: Gordon and Breach (1996). p.11-45

[5] R. Kretschmer and K.Binder, Phys. Rev. B. Vol.20, №3, 1065-1076 (1979).

[6] M.D. Glinchuk, E.A. Eliseev, V.A. Stephanovich. Physica B Vol. 332,- P.356-370 (2002).

[7] V. Fridkin, S. Ducharme, Ferroelectricity at the Nanoscale. Basics and Applications. Springer Heidelberg New York Dordrecht London. ISBN 978-3-642-41007-9 (2013). DOI 10.1007/978-3-642-41007-9

[8] R.V.Wang, D. D. Fong, F. Jiang, M. J. Highland, P. H. Fuoss, Carol Thompson, A. M. Kolpak, J. A. Eastman, S. K. Streiffer, A. M. Rappe, and G. B. Stephenson. Phys. Rev. Lett. 102, 047601 (2009).

[9] M.J. Highland, T.T. Fister, M.-I. Richard, D.D. Fong, P.H. Fuoss, C.Thompson, J.A. Eastman, S.K. Streiffer, and G.B. Stephenson. Phys. Rev. Lett. 105, 167601 (2010).

[10] M. J. Highland, T. T. Fister, D. D. Fong, P. H. Fuoss, Carol Thompson, J. A. Eastman, S. K. Streiffer, and G. B. Stephenson. Phys. Rev. Lett. **107**, 187602 (2011).

[11] G. B. Stephenson and M.J. Highland. Phys. Rev. B 84, 064107 (2011).

[12] Lars Onsager, Phys.Rev.**65**, 117, (1944)

[13] L.Landau and Lifschitz,Statistical Physics Part 1(Pergamon,Oxford,1980)

[14] Bune, Alexander V., Vladimir M. Fridkin, Stephen Ducharme, Lev M. Blinov, Serguei P. Palto, Alexander V. Sorokin, S. G. Yudin, and A. Zlatkin. *Nature* 391, no. 6670: 874-877 (1998).

[15] D. D. Fong, C. Cionca, Y. Yacoby, G. B. Stephenson, J. A. Eastman, P. H. Fuoss, S. K. Streiffer, Carol Thompson, R. Clarke, R. Pindak, and E. A. Stern. *Physical Review B* 71, no. 14: 144112 (2005).

[16] André Chanthbouala, Vincent Garcia, Ryan O. Cherifi, Karim Bouzehouane, Stéphane Fusil, Xavier Moya, Stéphane Xavier Hiroyuki Yamada, Cyrile Deranlot, Neil D. Mathur, Manuel Bibes, Agnès Barthélémy, Julie Grollier. *Nature materials* 11, no. 10: 860-864 (2012)

[17] Pertsev, N. A., A. G. Zembilgotov, and A. K. Tagantsev. *Physical Review Letters* **80**, 1998 (1988)

[18] Javier Junquera, and Philippe Ghosez. *Nature* 422, no. 6931: 506-509 (2003).

[19] Chun-Gang Duan, Renat F. Sabirianov, Wai-Ning Mei, Sitaram S. Jaswal, and Evgeny Y. Tsymbal. *Nano letters* **6**, 483 (2006).

[20] G.M.Guro, I.I.Ivanchik, N.F. Kovtonuk. Fiz. Tverd. Tela 10, 134-143 (1968) [Sov. Phys.-Solid State 10 100-109 (1968)]

[21] E. V. Chenskii, Fiz. Tverd. Tela (Leningrad) **12**, 586 (1970).

[22] I. I. Ivanchik, Ferroelectrics, 145, 149-61 (1993).

[23] Y. Watanabe, J. Appl. Phys., 83, 2179 -93, (1998).

[24] V.M. Fridkin, Ferroelectrics semiconductors, Consultant Bureau, New-York and London (1980).

[25] S.V. Kalinin, and D.A. Bonnell. Physical Review B 63, no. 12: 125411 (2001).

[26] S.V. Kalinin, Matthew R. Suchomel, Peter K. Davies, and Dawn A. Bonnell. Journal of the American Ceramic Society 85, no. 12: 3011-3017 (2002).

[27] N. Balke, B. Winchester, Wei Ren, Y.H. Chu, A.N. Morozovska, E.A. Eliseev, M. Huijben, R.K. Vasudevan, P.Maksymovych, J.Britson, S.Jesse, I.Kornev, R. Ramesh, L. Bellaiche, L. Q. Chen, and S.V. Kalinin. Nature Physics **8**, 81–88 (2012)

[28] P. Paruch, T. Tybell, J.-M. Triscone. Applied Physics Letters 79, no. 4: 530-532 (2001).

[29] S.V. Kalinin, B.J. Rodriguez, S. Jesse, E. Karapetian, B. Mirman, E.A. Eliseev, and A.N. Morozovska. Annu. Rev. Mater. Res. 37: 189-238 (2007).

[30] P. Maksymovych, A.N. Morozovska, Pu Yu, E.A. Eliseev, Y.-H. Chu, R. Ramesh, A.P. Baddorf, S.V. Kalinin. Nano Letters **12**, 209–213 (2012)

[31] A.K. Tagantsev, L.E. Cross, J. Fousek, Domains in Ferroic Crystals and Thin Films. Dordrecht : Springer, 2010. - 827 p.

[32] E.V. Chenskii, Fizika tverdogo tela 14, 2241 (1972).

[33] B.V. Selyuk, Kristallografiya 16, 356 (1971).

[34] B.M. Darinskii, A.P. Lazarev, A.S. Sidorkin, Kristallografiya 36, 757 (1991).

[35] Ch.-Lin Jia, V. Nagarajan, J.-Q. He, L. Houben, T. Zhao, R. Ramesh, K. Urban, and R. Waser. Nature materials 6, 64-69. (2007).

[36] A. K. Tagantsev, G. Gerra, and N. Setter. Phys. Rev. B **77**, 174111 (2008).

[37] I.I. Ivanchik, Fiz. Tverd. Tela **3**, 3731 (1961) [Sov. Phys. Solid State **3**, 2705 (1962)

[38] Y. L. Li, S. Y. Hu, Z. K. Liu, and L. Q. Chen. *Applied physics letters* 81, no. 3: 427-429 (2002).

[39] Sergei V. Kalinin, Edgar Karapetian, and Mark Kachanov. *Physical Review B* 70, 184101 (2004).

[40] E.A. Eliseev, A.N. Morozovska, G.S. Svechnikov, Venkatraman Gopalan, and V. Ya. Shur, // Static conductivity of charged domain walls in uniaxial ferroelectric semiconductors / Phys. Rev. B 83, 235313 (2011).

[41] M. Y. Gureev, A. K. Tagantsev, and N. Setter, Physical Review B 83, 184104, (2011).

[42] E.A. Eliseev, A.N. Morozovska, G.S. Svechnikov, Peter Maksymovych, S.V. Kalinin. Domain wall conduction in multiaxial ferroelectrics: impact of the wall tilt, curvature, flexoelectric coupling, electrostriction, proximity and finite size effects. Phys. Rev.B. **85**, 045312 (2012)

[43] S. M. Sze, Physics of Semiconductor Devices, 2nd ed. (Wiley-Interscience, New York, 1981).

[44] N.W. Ashcroft, N.D. Mermin, *Solid state physics* (Holt, Rinehart and Winston, New York, 1976) - 826 pages.

[45] S. M. Sze, *Physics of Semiconductor Devices*, 2nd ed. (Wiley-Interscience, New York, 1981).

[46] A.I. Anselm, *Introduction to semiconductor theory* (Mir, Moscow, Prentice-Hall, Englewood Cliffs, NJ, 1981).

[47] See Appendix S2 in supplementary material as [URL will be inserted by AIP] for details of calculation.

[48] J. X. Zhang, Y. L. Li, Y. Wang, Z. K. Liu, L. Q. Chen, Y. H. Chu, F. Zavaliche, and R. Ramesh. J. Appl. Phys. 101, 114105 (2007).

[49] M.J. Haun, E. Furman, S.J. Jang, and L.E. Cross, Ferroelectrics **99**, 63 (1989).

**Supplemental Materials to**

**Finite size effects in ferroelectric-semiconductor thin films under open-circuited electric boundary conditions**

Eugene A. Eliseev[1], Sergei V. Kalinin[2], Anna N. Morozovska[3*]

[1]Institute of Problems for Material Sciences, NAS of Ukraine, 03028 Kiev, Ukraine

[2]The Center for Nanophase Materials Sciences,
Oak Ridge National Laboratory, Oak Ridge, TN 37922

[3]Institute of Physics, NAS of Ukraine, 03028 Kiev, Ukraine

---

[*] Corresponding author: anna.n.morozovska@gmail.com

# Appendix S1. Debye screening radius and critical concentrations

**Figures S1a,b** illustrate the dependence of Debye screening radius vs. carriers concentration and temperature. Note, that the temperature dependence of Debye screening radius is almost indifferent on ferroelectric material parameters, it depends on $n_0$ and $\varepsilon_{33}^b$ only. **Figures S1c,d** illustrate the contour maps of the critical concentration $n_{cr}$ and $n_0^{cr}$ in coordinates "temperature $T$ – film thickness $L$"

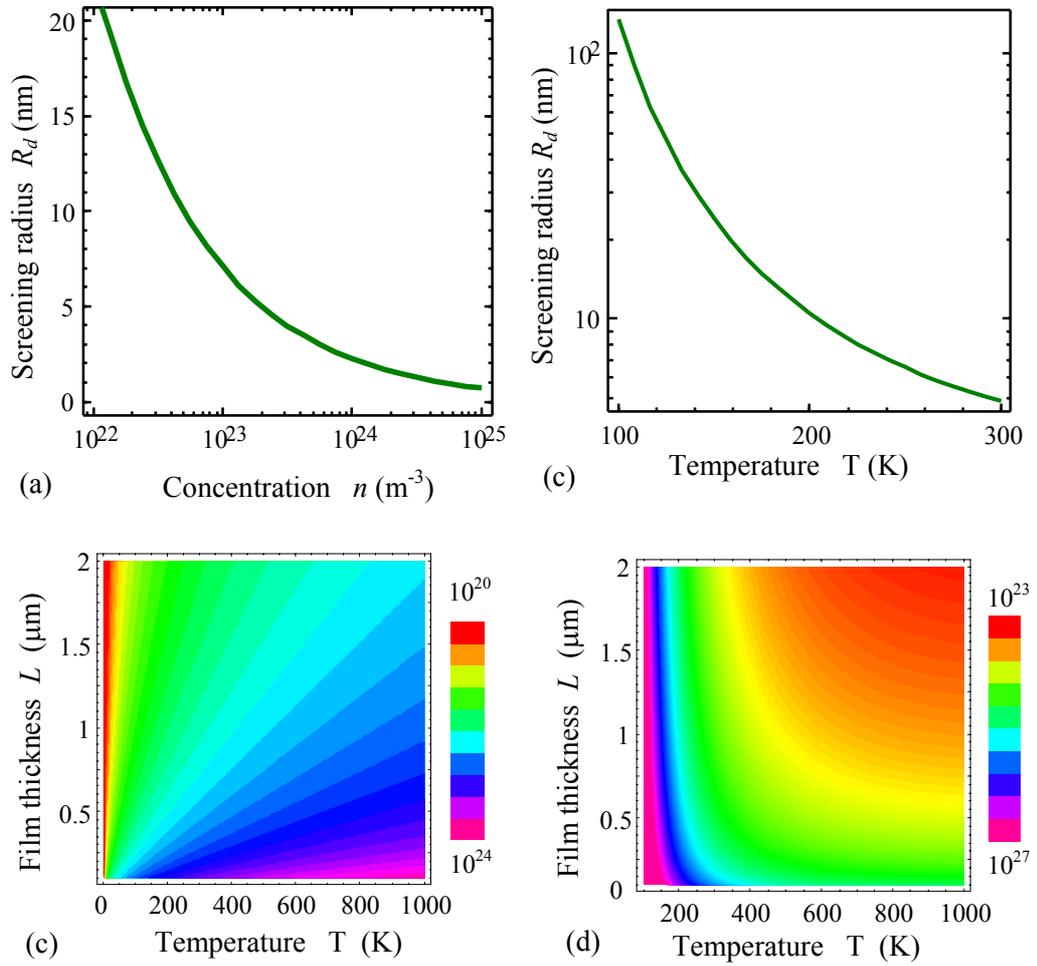

**Figure S1. (a)** Debye screening radius vs. carriers concentration calculated at 293°K. **(b)** Debye screening radius vs. temperature calculated at $n_0 = 10^{25}$m$^{-3}$, $E_d = 0.1$ eV and $\varepsilon_{33}^b = 7$. Contour maps of the critical concentration $n_{cr}$ **(c)** and $n_0^{cr}$ **(d)** in coordinates "temperature $T$ – film thickness $L$" calculated for $\varepsilon_{33}^b = 7$, $E_d = 0.1$ eV and $g = 10^{-10}$ C$^{-2}$m$^3$J. Color scale is the carrier concentration in m$^{-3}$.

# Appendix S1. Debye screening radius and critical concentrations

**Figures S1a,b** illustrate the dependence of Debye screening radius vs. carriers concentration and temperature. Note, that the temperature dependence of Debye screening radius is almost indifferent on ferroelectric material parameters, it depends on $n_0$ and $\varepsilon_{33}^b$ only. **Figures S1c,d** illustrate the contour maps of the critical concentration $n_{cr}$ and $n_0^{cr}$ in coordinates "temperature $T$ – film thickness $L$"

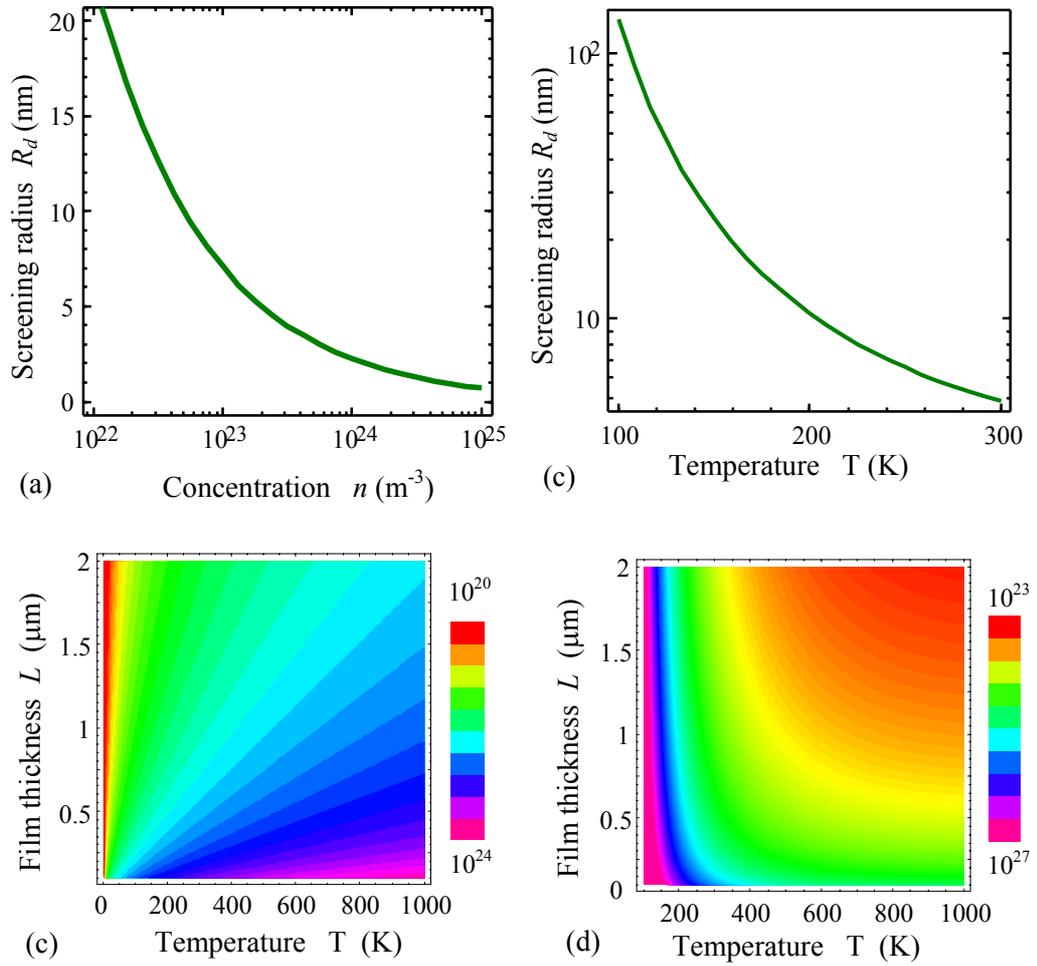

**Figure S1. (a)** Debye screening radius vs. carriers concentration calculated at 293°K. **(b)** Debye screening radius vs. temperature calculated at $n_0 = 10^{25}$m$^{-3}$, $E_d = 0.1$ eV and $\varepsilon_{33}^b = 7$. Contour maps of the critical concentration $n_{cr}$ **(c)** and $n_0^{cr}$ **(d)** in coordinates "temperature $T$ – film thickness $L$" calculated for $\varepsilon_{33}^b = 7$, $E_d = 0.1$ eV and $g = 10^{-10}$ C$^{-2}$m$^3$J. Color scale is the carrier concentration in m$^{-3}$.

**Appendix S2. Stability of paraelectric phase with respect to homogeneous fluctuations**

Let us write the linearized system of equation in the form

$$\begin{cases} \alpha P_3 - g\dfrac{\partial^2 P_3}{\partial z^2} = -\dfrac{\partial \varphi}{\partial z}, \\ \dfrac{\partial^2 \varphi}{\partial z^2} - \dfrac{\varphi}{R_d^2} = \dfrac{1}{\varepsilon_{33}^b \varepsilon_0}\dfrac{\partial P_3}{\partial z} \end{cases}$$

(S1.1)

with appropriate boundary conditions:

$$\begin{cases} \left(P_3 \pm \lambda \dfrac{\partial P_3}{\partial z}\right)\Big|_{\pm L/2} = 0, \\ \left(\varepsilon_0 \varepsilon_{33}^b \dfrac{\partial \varphi}{\partial z} - P_3\right)\Big|_{\pm L/2} = 0. \end{cases}$$

(S1.2)

Differentiation of the second of the equations (S1.1) gives

$$\dfrac{\partial^3 \varphi}{\partial z^3} - \dfrac{1}{R_d^2}\dfrac{\partial \varphi}{\partial z} = \dfrac{1}{\varepsilon_{33}^b \varepsilon_0}\dfrac{\partial^2 P_3}{\partial z^2}.$$

(S1.3a)

At the same time we could get from the first of the equations (S1.1) the following relations

$$\dfrac{\partial \varphi}{\partial z} = -\alpha P_3 + g\dfrac{\partial^2 P_3}{\partial z^2} \quad \text{and} \quad \dfrac{\partial^3 \varphi}{\partial z^3} = -\alpha \dfrac{\partial^2 P_3}{\partial z^2} + g\dfrac{\partial^4 P_3}{\partial z^4}.$$

(S1.3b)

Inserting relations (S1.3b) into (S1.3a), we could formally exclude electrostatic potential and get one equation for potential in the following form

$$g\dfrac{\partial^4 P_3}{\partial z^4} - \left(\alpha + \dfrac{g}{R_d^2} + \dfrac{1}{\varepsilon_{33}^b \varepsilon_0}\right)\dfrac{\partial^2 P_3}{\partial z^2} + \dfrac{\alpha}{R_d^2}P_3 = 0.$$

(S1.4)

In the same way, we could get boundary conditions for (S1.4) from (S1.2) as follows

$$\begin{cases} \left(P_3 \pm \lambda \dfrac{\partial P_3}{\partial z}\right)\Big|_{\pm L/2} = 0, \\ \left(g\dfrac{\partial^2 P_3}{\partial z^2} - \alpha P_3 - \dfrac{P_3}{\varepsilon_0 \varepsilon_{33}^b}\right)\Big|_{\pm L/2} = 0. \end{cases}$$

(S1.5)

Let us look for the solution of (S1.4) in the form $P_3 \sim \exp(wz)$, where inverse characteristic length $w$ satisfies the following equation:

$$g w^4 - \left(\alpha + \dfrac{g}{R_d^2} + \dfrac{1}{\varepsilon_{33}^b \varepsilon_0}\right)w^2 + \dfrac{\alpha}{R_d^2} = 0$$

(S1.6a)

Its solutions could be written as

$$w_{1,2}^2 = \dfrac{1}{2}\left(\dfrac{\alpha}{g} + \dfrac{1}{R_d^2} + \dfrac{1}{\varepsilon_{33}^b \varepsilon_0 g} \pm \sqrt{\left(\dfrac{\alpha}{g} + \dfrac{1}{R_d^2} + \dfrac{1}{\varepsilon_{33}^b \varepsilon_0 g}\right)^2 - \dfrac{4\alpha}{g R_d^2}}\right)$$

(S1.6b)

It is seen that at $\alpha < 0$ one of the roots of biquadratic equations (S1.6a) is always negative, while the other is positive, which means that one of the $w$ values is purely imaginary. Note, that in most cases $\varepsilon_{33}^b \varepsilon_0 g \ll \{R_d^2, g/|\alpha|\}$, hence the following approximations are valid

$$w_1 \approx \frac{\sqrt{\varepsilon_{33}^b \varepsilon_0 \alpha}}{R_d}, \quad w_2 \approx \frac{1}{\sqrt{\varepsilon_{33}^b \varepsilon_0 g}} \quad \text{at} \quad \varepsilon_{33}^b \varepsilon_0 g \ll \{R_d^2, g/|\alpha|\} \tag{S1.6c}$$

Now we could write the general solution of Eq.(S1.4) in the form:

$$P_3 = s_1 \sinh(w_1 z) + s_2 \sinh(w_2 z) + c_1 \cosh(w_1 z) + c_2 \cosh(w_2 z) \tag{S1.7}$$

The four constants $s_i$ and $c_i$ should be found from boundary conditions (S1.5). Formal solution is zero, since we have the system of homogeneous linear equations for $s_i$ and $c_i$, but we are interested in the stability analysis, hence we should look for zero point of the corresponding determinant of the linear equations system for $s_i$ and $c_i$. In the case of Eqs. (S1.5) and (S1.7) one could see that general 4×4 matrix is split on two independent diagonal blocks 2×2, which gives the following equation for instability point:

$$\begin{aligned}&\left(\left(w_1 \lambda \cosh\left(\frac{w_1 L}{2}\right) + \sinh\left(\frac{w_1 L}{2}\right)\right)\left(\frac{\alpha}{g} + \frac{1}{\varepsilon_{33}^b \varepsilon_0 g} - w_2^2\right)\sinh\left(\frac{w_2 L}{2}\right) - \right.\\ &\left.\left(w_2 \lambda \cosh\left(\frac{w_2 L}{2}\right) + \sinh\left(\frac{w_2 L}{2}\right)\right)\left(\frac{\alpha}{g} + \frac{1}{\varepsilon_{33}^b \varepsilon_0 g} - w_1^2\right)\sinh\left(\frac{w_1 L}{2}\right)\right) \\ &\times \left(\left(\cosh\left(\frac{w_1 L}{2}\right) + w_1 \lambda \sinh\left(\frac{w_1 L}{2}\right)\right)\left(\frac{\alpha}{g} + \frac{1}{\varepsilon_{33}^b \varepsilon_0 g} - w_2^2\right)\cosh\left(\frac{w_2 L}{2}\right) - \right.\\ &\left.\left(\cosh\left(\frac{w_2 L}{2}\right) + w_2 \lambda \sinh\left(\frac{w_2 L}{2}\right)\right)\left(\frac{\alpha}{g} + \frac{1}{\varepsilon_{33}^b \varepsilon_0 g} - w_1^2\right)\cosh\left(\frac{w_1 L}{2}\right)\right) = 0\end{aligned} \tag{S1.8a}$$

Here the first and the second factors correspond to spatially antisymmetric ($s_1 \sinh(w_1 z) + s_2 \sinh(w_2 z)$) and symmetric ($c_1 \cosh(w_1 z) + c_2 \cosh(w_2 z)$) solutions respectively. If the possible inhomogeneity of the equations (S1.1)-(S1.2) could be ascribed to one of these types, then these types of the solutions could be considered separately.

Let us start with spatially symmetric solutions. For the case of $\varepsilon_{33}^b \varepsilon_0 g \gg \{R_d^2, g/|\alpha|\}$ (see Eqs.(S1.6c)) and taking into account that $L \gg \sqrt{\varepsilon_{33}^b \varepsilon_0 g}$ (i.e. $w_2 L \gg 1$) one could get denoting $1/w_2$ as $\xi = \sqrt{\varepsilon_{33}^b \varepsilon_0 g}$

$$\cosh\left(\frac{w_1 L}{2}\right)\left(\left(w_1^2 - \frac{1}{\xi^2}\right)\left(1 + \frac{\lambda}{\xi}\right) - \frac{\alpha}{g}\frac{\lambda}{\xi}\right) + \lambda w_1 \sinh\left(\frac{w_1 L}{2}\right)\frac{\alpha}{g} \approx 0 \tag{S1.8b}$$

**Appendix S2. Stability of paraelectric phase with respect to homogeneous fluctuations**

Let us write the linearized system of equation in the form for ferroelectric media

$$\alpha P_3 - g_{33}\frac{\partial^2 P_3}{\partial z^2} - g_{55}\frac{\partial^2 P_3}{\partial x^2} = -\frac{\partial \varphi}{\partial z} \qquad (S2.1a)$$

$$\varepsilon_{33}^b \frac{\partial^2 \varphi^{(in)}}{\partial z^2} + \varepsilon_{11}^b \frac{\partial^2 \varphi^{(in)}}{\partial x^2} - \varepsilon_{33}^b \frac{\varphi^{(in)}}{R_d^2} = \frac{1}{\varepsilon_0}\frac{\partial P_3}{\partial z} \qquad (S2.1b)$$

and for vacuum (air) outside the system

$$\frac{\partial^2 \varphi^{(out)}}{\partial z^2} + \frac{\partial^2 \varphi^{(out)}}{\partial x^2} = 0 \qquad (S2.1c)$$

with appropriate boundary conditions:

$$\left(P_3 \pm \lambda \frac{\partial P_3}{\partial z}\right)\bigg|_{\pm L/2} = 0, \qquad (S2.2a)$$

$$\left(-\varepsilon_0 \varepsilon_{33}^b \frac{\partial \varphi^{(in)}}{\partial z} + P_3 + \varepsilon_0 \frac{\partial \varphi^{(out)}}{\partial z}\right)\bigg|_{\pm L/2} = 0. \qquad (S2.2b)$$

$$\left(\varphi^{(out)} - \varphi^{(in)}\right)\bigg|_{\pm L/2} = 0. \qquad (S2.2c)$$

Let us consider harmonic like fluctuations

$$P_3 = P_k(z)\exp(i k x), \quad \varphi^{(in)} = \varphi_k^{(in)}(z)\exp(i k x), \quad \varphi^{(out)} = \varphi_k^{(out)}(z)\exp(i k x)$$

Equations for amplitudes

$$(\alpha + g_{55}k^2)P_k - g_{33}\frac{\partial^2 P_k}{\partial z^2} = -\frac{\partial \varphi_k}{\partial z} \qquad (S2.3a)$$

$$\frac{\partial^2 \varphi_k^{(in)}}{\partial z^2} - \kappa^2 \varphi_k^{(in)} = \frac{1}{\varepsilon_0 \varepsilon_{33}^b}\frac{\partial P_k}{\partial z} \qquad (S2.3b)$$

$$\frac{\partial^2 \varphi_k^{(out)}}{\partial z^2} - k^2 \varphi_k^{(out)} = 0 \qquad (S2.3c)$$

Here we introduced parameter

$$\kappa^2 = \frac{\varepsilon_{11}^b}{\varepsilon_{33}^b}k^2 + \frac{1}{R_d^2}$$

Differentiation of the second of the equations (S2.3a) and (S2.3b) gives

$$\left(\frac{\partial^2}{\partial z^2} - \kappa^2\right)\left[(\alpha + g_{55}k^2)P_k - g_{33}\frac{\partial^2 P_k}{\partial z^2}\right] = \left(\frac{\partial^2}{\partial z^2} - \kappa^2\right)\left[-\frac{\partial \varphi_k}{\partial z}\right]. \qquad (S2.4a)$$

$$\frac{\partial}{\partial z}\left(\frac{\partial^2}{\partial z^2} - \kappa^2\right)\varphi_k^{(in)} = \frac{1}{\varepsilon_0 \varepsilon_{33}^b}\frac{\partial^2 P_k}{\partial z^2} \qquad (S2.4b)$$

Hence, one could exclude potential amplitude from Eq. (S2.4a) and get single equation for polarization amplitude in the form:

$$\left(\frac{\partial^2}{\partial z^2} - \kappa^2\right)\left[\left(\alpha + g_{55}k^2\right)P_k - g_{33}\frac{\partial^2 P_k}{\partial z^2}\right] = -\frac{1}{\varepsilon_0 \varepsilon_{33}^b}\frac{\partial^2 P_k}{\partial z^2} \quad (S2.5a)$$

The differentiation of Eq.(S2.3a) with respect to z gives

$$\frac{\partial^2 \varphi_k}{\partial z^2} = -\left(\alpha + g_{55}k^2\right)\frac{\partial P_k}{\partial z} + g_{33}\frac{\partial^3 P_k}{\partial z^3}$$

After substituting this relation into Eq.(S2.3b) one could get the following relation for potential:

$$\varphi_k^{(in)} = \frac{1}{\kappa^2}\left(-\left(\alpha + g_{55}k^2 + \frac{1}{\varepsilon_0 \varepsilon_{33}^b}\right)\frac{\partial P_k}{\partial z} + g_{33}\frac{\partial^3 P_k}{\partial z^3}\right) \quad (S2.5b)$$

Let us look for the solution of (S2.5) in the form $P_3 \sim \exp(wz)$, where inverse characteristic length $w$ satisfies the following equation:

$$w^4 - \left(\frac{\alpha + g_{55}k^2}{g_{33}} + \kappa^2 + \frac{1}{\varepsilon_{33}^b \varepsilon_0 g_{33}}\right)w^2 + \frac{\alpha + g_{55}k^2}{g_{33}}\kappa^2 = 0 \quad (S2.6a)$$

Its solutions could be written as

$$w_{1,2}^2 = \frac{1}{2}\left(\frac{\alpha + g_{55}k^2}{g_{33}} + \kappa^2 + \frac{1}{\varepsilon_{33}^b \varepsilon_0 g_{33}} \pm \sqrt{\left(\frac{\alpha + g_{55}k^2}{g_{33}} + \kappa^2 + \frac{1}{\varepsilon_{33}^b \varepsilon_0 g_{33}}\right)^2 - 4\frac{\alpha + g_{55}k^2}{g_{33}}\kappa^2}\right) \quad (S2.6b)$$

It should be noted, that in most cases $\varepsilon_{33}^b \varepsilon_0 g_{33} \ll \{1/\kappa^2, g_{33}/|\alpha|, g_{55}/|\alpha|\}$, hence the following approximations are valid

$$w_1 \approx \sqrt{\frac{(\alpha + g_{55}k^2)\kappa^2}{\alpha + g_{55}k^2 + g_{33}\kappa^2 + \frac{1}{\varepsilon_{33}^b \varepsilon_0}}} \approx \sqrt{\varepsilon_{33}^b \varepsilon_0 (\alpha + g_{55}k^2)\kappa^2}, \quad w_2 \approx \frac{1}{\sqrt{\varepsilon_{33}^b \varepsilon_0 g_{33}}} \quad (S2.6c)$$

Now we could write the general solution of Eq.(S2.5) in the form:

$$P_3 = s_1 \sinh(w_1 z) + s_2 \sinh(w_2 z) + c_1 \cosh(w_1 z) + c_2 \cosh(w_2 z) \quad (S2.7a)$$

The four constants $s_i$ and $c_i$ should be found from boundary conditions (S2.2). Formal solution is zero, since we have the system of homogeneous linear equations for $s_i$ and $c_i$, but we are interested in the stability analysis, hence we should look for zero point of the corresponding determinant of the linear equations system for $s_i$ and $c_i$. One could show that symmetric and asymmetric solutions could be considered separately.

$$P_3 = c_1 \cosh(w_1 z) + c_2 \cosh(w_2 z) \quad (S2.8a)$$

Thus, the electrostatic potential could be found from Eqs.(S2.5b) and (S2.8a) as

$$\varphi_k^{(in)} = \frac{1}{\kappa^2}\left(-\left(\alpha + g_{55}k^2 + \frac{1}{\varepsilon_0\varepsilon_{33}^b}\right)w_1 + g_{33}w_1^3\right)c_1\sinh(w_1 z) +$$
$$+ \frac{1}{\kappa^2}\left(-\left(\alpha + g_{55}k^2 + \frac{1}{\varepsilon_0\varepsilon_{33}^b}\right)w_2 + g_{33}w_2^3\right)c_2\sinh(w_2 z) \quad \text{(S2.8b)}$$

Expression for the electric filed could be easily found from Eq.(S2.8b) as

$$\frac{\partial\varphi_k^{(in)}}{\partial z} = \frac{w_1^2}{\kappa^2}\left(-\left(\alpha + g_{55}k^2 + \frac{1}{\varepsilon_0\varepsilon_{33}^b}\right) + g_{33}w_1^2\right)c_1\cosh(w_1 z) +$$
$$+ \frac{w_2^2}{\kappa^2}\left(-\left(\alpha + g_{55}k^2 + \frac{1}{\varepsilon_0\varepsilon_{33}^b}\right) + g_{33}w_2^2\right)c_2\cosh(w_2 z) \quad \text{(S2.8c)}$$

The solution of Eq. (S2.3c), decaying at $z \to \pm\infty$ is

$$\varphi_k^{(out)} = \pm f\exp\left(\mp |k|\left(z \pm \frac{L}{2}\right)\right) \quad \text{(S2.8d)}$$

It is seen from Eqs.(S2.8a, b, d) that one has three constants to be determined from the boundary conditions (S2.2). Hence the substitution of Eqs.(S2.8a, b, d) into Eqs.(S2.2) gives the following system of equations

$$c_1\left(\cosh\left(\frac{w_1 L}{2}\right) + \lambda w_1\sinh\left(\frac{w_1 L}{2}\right)\right) + c_2\left(\cosh\left(\frac{w_2 L}{2}\right) + \lambda w_2\sinh\left(\frac{w_2 L}{2}\right)\right) = 0, \quad \text{(S2.9a)}$$

$$\left(1 - \varepsilon_0\varepsilon_{33}^b\frac{w_1^2}{\kappa^2}\left(-\left(\alpha + g_{55}k^2 + \frac{1}{\varepsilon_0\varepsilon_{33}^b}\right) + g_{33}w_1^2\right)\right)c_1\cosh\left(\frac{w_1 L}{2}\right) +$$
$$\left(1 - \varepsilon_0\varepsilon_{33}^b\frac{w_2^2}{\kappa^2}\left(-\left(\alpha + g_{55}k^2 + \frac{1}{\varepsilon_0\varepsilon_{33}^b}\right) + g_{33}w_2^2\right)\right)c_2\cosh\left(\frac{w_2 L}{2}\right) - |k|\varepsilon_0 f = 0. \quad \text{(S2.9b)}$$

$$\frac{1}{\kappa^2}\left(-\left(\alpha + g_{55}k^2 + \frac{1}{\varepsilon_0\varepsilon_{33}^b}\right)w_1 + g_{33}w_1^3\right)c_1\sinh\left(\frac{w_1 L}{2}\right) +$$
$$+ \frac{1}{\kappa^2}\left(-\left(\alpha + g_{55}k^2 + \frac{1}{\varepsilon_0\varepsilon_{33}^b}\right)w_2 + g_{33}w_2^3\right)c_2\sinh\left(\frac{w_2 L}{2}\right) - f = 0. \quad \text{(S2.9c)}$$

After exclusion of constant $f$ one get the following system

$$c_1\left(\cosh\left(\frac{w_1 L}{2}\right) + \lambda w_1\sinh\left(\frac{w_1 L}{2}\right)\right) + c_2\left(\cosh\left(\frac{w_2 L}{2}\right) + \lambda w_2\sinh\left(\frac{w_2 L}{2}\right)\right) = 0, \quad \text{(S2.10a)}$$

$$\left(\cosh\left(\frac{w_1 L}{2}\right)+\varepsilon_0\left(\varepsilon_{33}^b \frac{w_1^2}{\kappa^2}\cosh\left(\frac{w_1 L}{2}\right)+|k|\frac{w_1}{\kappa^2}\sinh\left(\frac{w_1 L}{2}\right)\right)\left(\alpha+g_{55}k^2+\frac{1}{\varepsilon_0\varepsilon_{33}^b}-g_{33}w_1^2\right)\right)c_1+$$

$$+$$

$$\left(\cosh\left(\frac{w_2 L}{2}\right)+\varepsilon_0\left(\varepsilon_{33}^b \frac{w_2^2}{\kappa^2}\cosh\left(\frac{w_2 L}{2}\right)+|k|\frac{w_2}{\kappa^2}\sinh\left(\frac{w_2 L}{2}\right)\right)\left(\alpha+g_{55}k^2+\frac{1}{\varepsilon_0\varepsilon_{33}^b}-g_{33}w_2^2\right)\right)c_2=0.$$

(S2.10b)

It determinant should be zero, which gives the condition of the phase transition in the form

$$\left(\cosh\left(\frac{w_1 L}{2}\right)+\varepsilon_0\left(\varepsilon_{33}^b \frac{w_1^2}{\kappa^2}\cosh\left(\frac{w_1 L}{2}\right)+|k|\frac{w_1}{\kappa^2}\sinh\left(\frac{w_1 L}{2}\right)\right)\left(\alpha+g_{55}k^2+\frac{1}{\varepsilon_0\varepsilon_{33}^b}-g_{33}w_1^2\right)\right)$$

$$\times\left(\cosh\left(\frac{w_2 L}{2}\right)+\lambda w_2 \sinh\left(\frac{w_2 L}{2}\right)\right)=$$

$$=\left(\cosh\left(\frac{w_2 L}{2}\right)+\varepsilon_0\left(\varepsilon_{33}^b \frac{w_2^2}{\kappa^2}\cosh\left(\frac{w_2 L}{2}\right)+|k|\frac{w_2}{\kappa^2}\sinh\left(\frac{w_2 L}{2}\right)\right)\left(\alpha+g_{55}k^2+\frac{1}{\varepsilon_0\varepsilon_{33}^b}-g_{33}w_2^2\right)\right)\times$$

$$\times\left(\cosh\left(\frac{w_1 L}{2}\right)+\lambda w_1 \sinh\left(\frac{w_1 L}{2}\right)\right)$$

(S2.11a)

In the limit $w_2 L \gg 1$ one could rewrite (S2.11a) as

$$\left(\kappa^2\cosh\left(\frac{w_1 L}{2}\right)+\varepsilon_0\left(\varepsilon_{33}^b w_1^2\cosh\left(\frac{w_1 L}{2}\right)+|k|w_1\sinh\left(\frac{w_1 L}{2}\right)\right)\left(\alpha+g_{55}k^2+\frac{1}{\varepsilon_0\varepsilon_{33}^b}-g_{33}w_1^2\right)\right)(1+\lambda w_2)-$$

$$-\left(\kappa^2+\varepsilon_0\left(\varepsilon_{33}^b w_2^2+|k|w_2\right)\left(\alpha+g_{55}k^2+\frac{1}{\varepsilon_0\varepsilon_{33}^b}-g_{33}w_2^2\right)\right)\left(\cosh\left(\frac{w_1 L}{2}\right)+\lambda w_1 \sinh\left(\frac{w_1 L}{2}\right)\right)\approx 0$$

(S2.11b)

Using the following relation

$$\left(\alpha+g_{55}k^2+\frac{1}{\varepsilon_{33}^b\varepsilon_0}-g_{33}w^2\right)w^2=g_{33}\kappa^2\left(w^2+\frac{\alpha+g_{55}k^2}{g_{33}}\right)$$

Coming from Eq. (S2.6a), we rewrote (S2.11b) as

$$\left(\kappa^2\cosh\left(\frac{w_1 L}{2}\right)+\varepsilon_0\left(\varepsilon_{33}^b w_1^2\cosh\left(\frac{w_1 L}{2}\right)+|k|w_1\sinh\left(\frac{w_1 L}{2}\right)\right)\frac{g_{33}\kappa^2}{w_1^2}\left(w_1^2+\frac{\alpha+g_{55}k^2}{g_{33}}\right)\right)(1+\lambda w_2)-$$

$$-\left(\kappa^2+\varepsilon_0\left(\varepsilon_{33}^b w_2^2+|k|w_2\right)\frac{g_{33}\kappa^2}{w_2^2}\left(w_2^2+\frac{\alpha+g_{55}k^2}{g_{33}}\right)\right)\left(\cosh\left(\frac{w_1 L}{2}\right)+\lambda w_1 \sinh\left(\frac{w_1 L}{2}\right)\right)\approx 0$$

(S2.11c)

Finally, using a reasonable approximation for the characteristic root $w_2\approx(\varepsilon_{33}^b\varepsilon_0 g_{33})^{-1/2}$ and the fact that length scale $\sqrt{\varepsilon_0\varepsilon_{33}^b g_{33}}$ is usually smaller or even much smaller than lattice constant one could reach further simplification of Eq.(S2.11) in the form

$$\left(1+\varepsilon_0\varepsilon_{33}^b g_{33}\left(w_1^2+\frac{\alpha+g_{55}k^2}{g_{33}}\right)-\frac{2}{1+\lambda w_2}\right)\cosh\left(\frac{w_1 L}{2}\right)+$$
$$+\left(\frac{\varepsilon_0 g_{33}|k|}{w_1}\left(w_1^2+\frac{\alpha+g_{55}k^2}{g_{33}}\right)-\frac{2\lambda w_1}{1+\lambda w_2}\right)\sinh\left(\frac{w_1 L}{2}\right)\approx 0 \quad \text{(S2.11d)}$$

One could easily see that

$$\varepsilon_0\varepsilon_{33}^b g_{33}\left(w_1^2+\frac{\alpha+g_{55}k^2}{g_{33}}\right)<<1 \text{ and } \frac{\varepsilon_0 g_{33}|k|}{|w_1|}\left(w_1^2+\frac{\alpha+g_{55}k^2}{g_{33}}\right)<<1$$

in most cases.

Hence the condition (S2.11) is reduced to $\left(1-\frac{2}{1+\lambda w_2}\right)\cosh\left(\frac{w_1 L}{2}\right)-\frac{2\lambda w_1}{1+\lambda w_2}\sinh\left(\frac{w_1 L}{2}\right)\approx 0$

In the most of the cases the second term is much smaller than the first one since $|w_1|<<w_2$.

$$\left(1-\frac{2}{1+\lambda w_2}\right)\cosh\left(\frac{w_1 L}{2}\right)\approx 0 \quad \text{(S2.11e)}$$

Taking into account that $w_1\approx\sqrt{\varepsilon_{33}^b\varepsilon_0\left(\alpha+g_{55}k^2\right)\left(\frac{\varepsilon_{11}^b}{\varepsilon_{33}^b}k^2+\frac{1}{R_d^2}\right)}$ one could easily get the smallest root of Eq.(S2.11e) as

$$L\approx\frac{\pi}{\sqrt{\varepsilon_{33}^b\varepsilon_0\left(-\alpha-g_{55}k^2\right)\left(\frac{\varepsilon_{11}^b}{\varepsilon_{33}^b}k^2+\frac{1}{R_d^2}\right)}} \quad \text{(S2.12)}$$

This expression for the critical thickness should be further minimized with respect to wave vector k.

$$L_{min}\approx\frac{2\pi}{\left(-\frac{\alpha}{g_{55}}+\frac{\varepsilon_{33}^b}{\varepsilon_{11}^b R_d^2}\right)\sqrt{\varepsilon_{33}^b\varepsilon_0 g_{55}}} \quad \text{(S2.13)}$$

$$k_{min}=\sqrt{-\frac{\alpha}{2g_{55}}-\frac{\varepsilon_{33}^b}{2\varepsilon_{11}^b R_d^2}} \quad \text{(S2.14)}$$

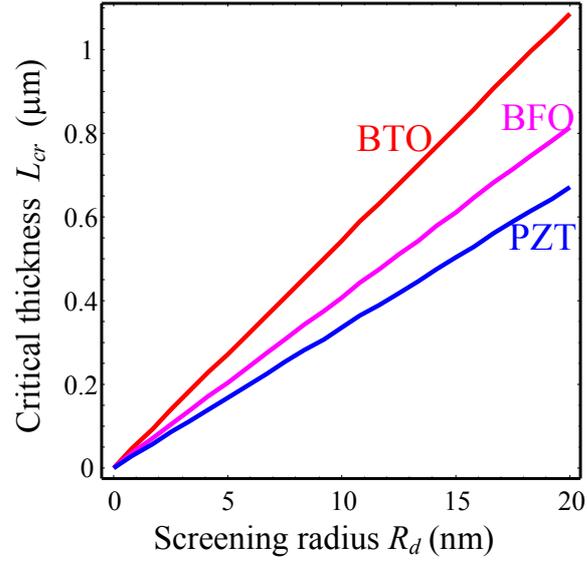

**Figure S2.** Film critical thickness vs. Debye screening radius calculated from Eq.(4b) at room temperature 293°K. Different curves correspond to different ferroelectric materials BiFeO$_3$ (BFO), BaTiO$_3$ (BTO), PbZr$_{40}$Ti$_{60}$O$_3$ (PZT), which material parameters are listed in the **Table 1**.